\documentstyle[twocolumn,epsbox]{jpsj} 
\def\runtitle{ 
Two-Band-Type Superconducting Instability in MgB$_2$
}
\def\runauthor{K. {\sc Yamaji}}
\title
{Two-Band-Type Superconducting Instability in MgB$_2$}
\author
{
Kunihiko {\sc Yamaji}\footnote{New address from April 1:
Nanoelectronics Research Institute, AIST Tsukuba Central 2, 
Tsukuba, Ibaraki 305-8568}
}
\inst
{
Electrotechnical Laboratory,
1-1-4 Umezono, Tsukuba, Ibaraki 305-8568; 
e-mail: yamaji-kuni@aist.go.jp
}
\recdate
{21 March 2001} 
\abst
{
Using the tight-binding method for the $\pi$-bands in MgB$_2$, 
the Hubbard on-site Coulomb interaction 
on two inequivalent boron $p_z$-orbitals is 
transformed into expressions in terms of $\pi$-band operators. 
For scattering processes relevant to the problem
in which a wave vector {\bf q} is parallel to $\hat{z}$, 
it is found to take a relatively simple form consisting of 
intra-band Coulomb scattering, interband pair scattering etc. 
with large constant coupling constants. 
This allows to get a simple expression
for the amplitude of interband pair scattering between two $\pi$-bands,
which diverges if the interband polarization function in it becomes
large enough.
The latter was approximately evaluated and found to be largely
enhanced in the band structure in MgB$_2$.
These results 
lead to a divergent interband pair scattering, meaning two-band-type 
superconducting instability with enhanced $T_c$. 
Adding a subsidiary BCS attractive interaction in each band 
into consideration,
a semi-quantitative gap equation is given, and $T_c$ and isotope exponent 
$\alpha$ are derived. The present instability is asserted 
to be the origin of high $T_c$ in MgB$_2$. 
}

\kword
{
MgB$_2$, interband polarization function, superconducting instability, 
two-band mechanism, interband pair transfer, 
Coulomb interaction, nesting, isotope effect
}

\begin{document}
\maketitle

The recent finding of 40K superconductivity in MgB$_2$ 
by the Akimitsu group\cite{rf:akimitsu} triggered enthusiastic 
studies both in experiment and theory to clarifiy the superconducting 
properties and mechanism of the new high-$T_c$ superconductor. 
Given $T_c \approx 40$K amply exceeding the upper bound of 
$T_c \sim 30$K which was estimated 
for conventional BCS-type superconductors,\cite{rf:mcmillan} 
it is a very important issue 
if its superconducting mechanism is conventional or not. 
Already Bud'ko et al.~\cite{rf:budko} has reported an 
isotope effect of $\alpha=0.26$ with respect of boron, 
indicating the importance of electron-phonon interaction. 
Band calculations have suggested that the dimensionless 
electron-phonon coupling constant 
$\lambda \sim 0.7$,\cite{rf:kortus,rf:pickett,rf:anderson}  
which would be 
able to give the $T_c$ value of the order of 40K if the 
Coulomb pseudo potential term $\mu^*$ is very small. 
Tunneling results indicate the s-wave nature of 
superconductivity.\cite{rf:tun1,rf:tun2,rf:tun3} 
All these suggests the BCS-type conventional superconductivity 
in MgB$_2$.

However, a detailed analysis of the specific heat indicated that 
the superconducting gap must be anisotropic or two-band-like.\cite{rf:junod} 
Non-conventional features were reported 
also about the power-law temperature dependence of 
$H_{c1}$\cite{rf:zhao} and the penetration depth.\cite{rf:zhao,rf:panago} 
Theoretically, Furukawa pointed out that the Fermi surfaces of
two $\pi$-bands of MgB$_2$ are close to perfect nesting,
which he suggests would give rise to antiferromagnetism;
the perfect nesting would be realized  
if holes did not enter lower $\sigma$-bands, 
breaking the equality of electron and hole numbers
in the $\pi$-bands.\cite{rf:furukawa} 
In this letter we report that the interband polarization function 
in the MgB$_2$ band even as it is
greatly enlarged for appropriate wave vectors 
between the two $\pi$-bands  
and that it can lead to 
a divergent enhancement of the amplitude 
of Coulomb-origin interband electron-pair transfer process. 
The latter means an instability to 
a two-band-type superconductivity.
If this mechanism is the 
dominant one, the type of superconductivity is s-wave-like 
in each band with each gap parameter of an sign opposite
between the two $\pi$-bands; 
the gap in the $\sigma$-bands are subsidiary and smaller
in a clear contrast to the electron-phonon scenario. 
This mechanism allows to qualitatively explain the isotope effect
when it takes account of the effect of the lattice subsidiarily.
According to this mechanism 
$T_c$ can go up further through improvement of the nesting. \


Our Hamiltonian is  a  Hubbard  model whose one-particle part 
is the tight-binding bands giving 
the two $\pi$-bands 
on the graphene-structure 
boron layers. 
Let us put two boron atoms in the unit cell taken by 
Wallace\cite{rf:wallace} at ($\pm a/2\surd{3}$, 0, 0), 
where $a$ is the distance between the equivalent nearest 
neighbor boron sites in the boron plane. Then the tight-binding 
bands of the $\pi$-bands are given by
\begin{eqnarray}
&\varepsilon&_{\pm}({\bf k}) =
\pm \vert h_{21}({\bf k}) \vert + 2t_z \cos(ck_z)  \nonumber \\ 
&=& \pm t [ 3 + 2\cos (ak_y) + 4\cos (\sqrt{3}ak_x / 2) 
\cos(ak_y / 2)]^{1/2}  \nonumber \\
& & + 2t_z \cos(ck_z), 
\label{eq:band}  
\end{eqnarray}
where
\begin{equation}
h_{21}({\bf k}) = t\{ 
1+\exp(-\mbox{i}{\bf a}_1 \cdot {\bf k}) + \exp[-\mbox{i}({\bf a}_1 
+{\bf a}_2) \cdot {\bf k}] \}
\label {eq:h21}
\end{equation}
with ${\bf a}_1= (\sqrt{3}a/2, a/2, 0)$ 
and ${\bf a}_2= (0, -a, 0)$; 
here $t$ and $t_z$ are the absolute value of transfer energy 
between the nearest neighbor boron $p_z$-orbitals in the plane 
and along the $z$-direction, respectively; $c$ is the 
$z$-lattice constant. 

The interaction part comes from the on-site Coulomb energy $U_0$ 
at each boron site. The annihilation operators at two inequivalent 
sites are Fourier-transformed and here denoted as
$a_{{\mib k}\sigma}$ and $b_{{\mib k}\sigma}$, 
where $\sigma$ means spin. 
When the one-electron part is diagonalized into two 
$\pi$-bands whose annihilation operators are 
$\alpha_{{\mib k}\sigma}$ and $\beta_{{\mib k}\sigma}$, 
the former operators can be linked to the latter 
through the following unitary transformation:
\begin{eqnarray}
a_{{\mib k}\sigma} &=& 
(1/\sqrt{2})\exp[-\mbox{i}\phi({\bf k})](\alpha_{{\mib k}\sigma}
-\beta_{{\mib k}\sigma}), \nonumber \\
b_{{\mib k}\sigma} &=& 
(1/\sqrt{2})\exp[\mbox{i}\phi({\bf k})](\alpha_{{\mib k}\sigma}  
+\beta_{{\mib k}\sigma}), 
\label{eq: trans} 
\end{eqnarray}
where $\phi({\bf k})$ is defined by half of the phase of 
$h_{21}({\bf k})$ in Eq. (\ref{eq:h21}).
If we restrict ${\bf q}$ 
appearing in the equation below to have only a $z$-component, 
this restricted part $H'_{\perp}$ of the interaction 
Hamiltonian is rewritten in terms of the new operators 
in a relatively simple form as follows:
\begin{eqnarray}
H'_{\perp} &=&
(1/N) \sum_{{\mib k_1},{\mib k_2},{\mib q} \perp \hat{x},\hat{y}} \nonumber \\
& &\{U[\alpha_{{\mib k}_1+{\mib q}\uparrow}^{\dag} \alpha_{{\mib k}_2+
{\mib q}\uparrow}\alpha_{{\mib k}_2 \downarrow}^{\dag} 
\alpha_{{\mib k}_1\downarrow}   
+ (\alpha \leftrightarrow \beta)] \nonumber \\
& &+U'[\alpha_{{\mib k}_1+{\mib q}\uparrow}^{\dag}\alpha_{{\mib k}_2
+{\mib q}\uparrow}\beta_{{\mib k}_2 \downarrow}^{\dag}
\beta_{{\mib k}_1\downarrow} 
+ (\alpha \leftrightarrow \beta)] \nonumber  \\
& &+K[\alpha_{{\mib k}_1+{\mib q}\uparrow}^{\dag}\beta_{{\mib k}_2
+{\mib q}\uparrow}\alpha_{{\mib k}_2 \downarrow}^{\dag}
\beta_{{\mib k}_1\downarrow} 
+ (\alpha \leftrightarrow \beta)]  \nonumber  \\
& &+L[\alpha_{{\mib k}_1+{\mib q}\uparrow}^{\dag}\beta_{{\mib k}_2
+{\mib q}\uparrow}\beta_{{\mib k}_2 \downarrow}^{\dag}
\alpha_{{\mib k}_1\downarrow} 
+ (\alpha \leftrightarrow \beta)]\},  
\label{eq:int}
\end{eqnarray}
with
\begin{equation}
U=U'=K=L=U_0/2,
\label{eq:coupling}
\end{equation}
where phase factors cancel each other 
since $\phi({\bf k})$ depends only 
on the in-plane component of {\bf k};
$(\alpha \leftrightarrow \beta)$ denotes a term identical to the 
preceding one except for exchange of $\alpha$ and $\beta$. 
The term with coupling constant $U$ gives rise to the intraband Coulomb
scattering, the term with $U'$ the interband Coulomb scattering,
the term with $K$ the interband electron-pair scattering 
and the term with $L$ the interband exchange scattering. 
This transformation is similar 
to that performed for the two-chain Hubbard model.\cite{rf:twochain} 
The interaction part with {\bf q} having non-$z$-component 
is very complicated.

\begin{figure}
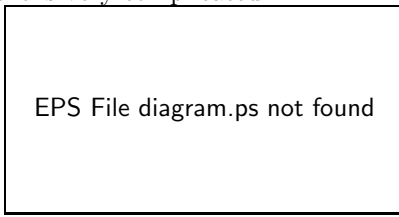

\begin{center}
\psbox[width=5cm]{diagram}
\caption{Ladder diagram enhancing the pair transfer process 
between two bands 1 and 2. Straight lines denote electron 
Green's function. Lateral dashed lines mean interactions 
with coupling constants $K$ and $U'$. 
Attached symbols 
illustrate an example among the diagrammatical processes.
The lower-case letter denotes a set of wave vector and frequency.} 
\label{fig:diagram}
\end{center}
\end{figure}

The interband pair-transfer term with $K$ in Eq. (\ref{eq:int}) 
is known to promote superconductivity.\cite{rf:kondo,rf:twochain} 
If $K$ is larger than $U$, 
the simple mean-field calculation gives a finite $T_c$.
The amplitude of interband pair-transfer process is known 
to be enhanced due to the higher 
order processes illustrated by the ladder-type diagram 
in Fig. \ref{fig:diagram}. 
When ${\bf q}$ in the diagram is chosen parallel to $\hat{z}$, 
the interaction denoted by a dotted line is given by
simple coupling constants $K$ or $U'$ and 
summation of the terms represented by this diagram gives the 
following expression:\cite{rf:yamaji}
\begin{equation}
\Gamma_{12} = K/[1 - (K+U')\Pi_{12}({\bf q})]
\label{eq:Kdiv}
\end{equation}
when the frequency transfer is set to zero for simplicity.
In this diagram with ${\bf q} \parallel \hat{z}$
the phase factor disappears.\

\begin{figure}
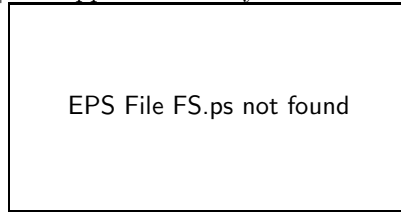

\begin{center}
\psbox[width=5cm]{FS}
\caption{Section of model Fermi surfaces with the $p_xk_z$-plane
in the case of $k_{z0} = \pi/2c$. 
Cones are considered to correctly reproduce the features of the Fermi surfaces 
in the neighborhood of the cone points,\cite{rf:kortus,rf:arm}
although the peripheral part
of the cones may not be a good approximation. Hatched part shows the 
electron Fermi surface. The diamond around the center is the 
hole Fermi surface.}
\label{fig:FS}
\end{center}
\end{figure}

Next, we first evaluate $\Pi_{12}({\bf q})$ in the case of the perfect
nesting and then discuss on it in the off-set case.
The $xy$-coordinates of the corner points, K (and also H), 
of the hexagonal Brillouin zone are 
$(\pm2\pi/\sqrt{3}a, \pm2\pi/3a)$ and $(0, \pm4\pi/3a)$.
At these points the intra-plane part of the band 
$\pm  \vert h_{21}({\bf k}) \vert $ in Eq. (\ref{eq:band})
vanishes. So for an appropriate value of Fermi energy $\varepsilon_F$ 
in the range between $-2t_z$ and $2t_z$, the Fermi surfaces 
for the $\pm$ bands, $i.e.$, electron and hole bands, respectively, 
cross the corner K-H-K line at points P$_{\pm}$ specified 
by $k_z = \pm \cos^{-1}(\varepsilon_F/2t_z) \equiv \pm k_{z0}$, 
repectively.
Then, the Fermi surfaces for the electron and hole bands  
in the neighborhood of P$_+$ are given by the 
$\pm$ branch of the following equation:
\begin{eqnarray}
\pm  \vert h_{21}({\bf k}) \vert
&=& 2t_z \cos(ck_{z0}) - 2t_z \cos(ck_z) \nonumber \\
&\cong& 2ct_z \sin (ck_{z0}) (k_z - k_{z0}).  
\label{eq:FS}
\end{eqnarray}
As seen in Fig. (\ref{fig:FS}), at least in the neighborhood of point P$_+$, 
where the second equality in Eqs. (\ref{eq:FS}) is justified, 
the electron and hole Fermi surfaces form the mirror images 
with respect of the $xy$-parallel plane 
crossing the corner line at P$_+$, as seen
in Fig. \ref{fig:FS} which is shown for another purpose.
There is another set of approximate mirror images around 
the P$_-$ point. Therefore, if the electron Fermi surface 
just below P$_-$ is translated by $2k_{z0}$ in the $k_z$-direction, 
it nests with the hole Fermi surface just 
below P$_+$.
Similarly, the electron Fermi surface part just 
above P$_+$ nests with the hole Fermi surface part 
just below P$_-$ if the former is translated by $-2k_{z0}$. 
In the neighborhood of the 
corner line K-H-K labeled by $(2\pi/\sqrt{3}a, 2\pi/3a)$ 
in the $k_xk_y$-coordinates,
expanding $h_{21}({\bf k})$ in powers of 
$p_x=k_x-2\pi/\sqrt{3}a$ and $p_y=k_y-2\pi/3a$, 
Eq. (\ref{eq:FS}) is approximated as 
\begin{equation}
\pm (\sqrt{3} /2) a\,t\,p_{\parallel}
\cong 2ct_z \sin (ck_{z0}) (k_z - k_{z0}), 
\label{eq:approxFS}
\end{equation}
where $p_{\parallel} \equiv \sqrt{p_x^2+p_y^2}$. 
This shows the Fermi surfaces around P$_+$ are 
approximated by cones as seen in Fig. \ref{fig:FS}. 

When $k_{z0} = \pi/2c$, one can confirm from 
Eq. (\ref{eq:FS}) that the electron and hole Fermi surfaces 
take identical forms and perfectly nest with each other 
by translation of $\pi/c$ 
exactly in the framework of the tight-binding bands
in accord with Ref.\cite{rf:furukawa}
The interband polarization function 
between the two $\pi$-bands
for zero frequency  is
defined by
\begin{equation}
\Pi_{12}({\bf Q}) =
\frac{1}{N} \sum_{\mib k} 
{f(\varepsilon_-({\bf k} + {\bf Q})) - f(\varepsilon_+({\bf k})) 
\over \varepsilon_+({\bf k}) - \varepsilon_-({\bf k} + {\bf Q})},
                             \label{eq:pi}
\end{equation}
where $N$ is the number of unit cells and $f(\varepsilon_{\pm})$ 
is the Fermi-Dirac distribution function. In order to capture 
gross features, we approximately evaluate this summation 
in the case of perfect nesting, approximating the right-hand 
side of Eq. (\ref{eq:band}) by terms linear in 
$p_{\parallel}$ and in $k_z-k_{z0}$ as
$\varepsilon_{\pm}({\bf k}) =
\pm (\sqrt{3} /2)\, a \, t\,p_{\parallel}
+ v_{\perp}(k_z - k_{z0})$
with $v_{\perp}=2ct_z\sin(ck_{z0})$ and $k_{z0} = \pi/2c$.
Then, the Fermi surfaces are given as in Fig. \ref{fig:FS}. 
The hole Fermi surface 
takes the form of two cones sticked at the bottoms with total height 
of $\pi/c$. It centers at $k_z=0$. The electron Fermi surface takes 
the identical form and centers at $k_z=\pi/c$ in the extended 
zone scheme. 
Since at each corner integration range is restricted to 120 degrees, 
we have two 360 degree integrals, counting six corners. 
The upper bound of 
$p_{\parallel}$ is set to $k_2 \equiv 2(\pi/\sqrt{3})^{1/2}/a$ 
so that the total integration area is equal to the area of the 
reduced Brillouin zone. 
For the wave vector ${\bf Q}_0 \equiv (0, 0, \pi/c)$ 
of perfect nesting, we gets
\begin{eqnarray}
\Pi_{12}({\bf Q} _0) &\cong&
\frac{V_{cell}}{\pi^2 v_{\parallel}} 
\{k_2\frac{\pi}{4c} - \frac{3}{2} \frac{v_{\perp}}{v_{\parallel}} 
(\frac{\pi}{2c})^2 - \frac{v_{\parallel}k_2^2}{2v_{\perp}} 
\ln(1-{v_{\perp}\frac{\pi}{2c} \over v_{\parallel}k_2})  \nonumber \\
&+&
\frac{1}{2} \frac{v_{\perp}}{v_{\parallel}} (\frac{\pi}{2c})^2 
\ln \frac{ 4\gamma^2 v_{\perp} \frac{\pi}{2c} v_{\parallel} 
(k_2 - \frac{v_{\perp}}{v_{\parallel}} \frac{\pi}{2c})}
{\pi^2 T^2} \}, 
 \label{eq:pi12} 
\end{eqnarray}
where $V_{cell} = \sqrt{3}a^2c/2$ is the volume of the unit cell, 
$v_{\parallel} = \sqrt{3}at/2$, 
$\gamma = 1.78107$ and $T$ is the temperature.
Putting $a=3.084 \AA$ and $c=3.522 \AA$ and using $t \sim 2.5 \mbox{eV}$
and $t_z \sim 1.5 \mbox{eV}$,\cite{rf:kortus} 
$\Pi_{12}({\bf Q}_0)$ is estimated as
\begin{equation}
\Pi_{12}({\bf Q}_0) \cong 0.053 + 0.139 \times \ln(30152/T) (\mbox{eV})^{-1},
 \label{eq: value} 
\end{equation}
where $T$ is in units of K. 
Putting $T=40$K, we get
$\Pi_{12}({\bf Q}_0) \cong 0.97$ (eV)$^{-1}$.
This value is much larger than the intra-band polarization 
function 
$\Pi_i(0) = V_{cell}v_{\perp}/4c^2v_{\parallel}^2 \cong 0.139$ (eV)$^{-1}$ 
of band $i=$1 or 2 for wave vector 
equal to zero.
For wave vector ${\bf Q}$ slightly deviating from 
${\bf Q}_0$, $\Pi_{12}({\bf Q})$
decreases in the way that $T^2$ in the argument of $\ln$ 
in Eq. (\ref{eq:pi12}) is replaced by 
$T^2 + c_{\parallel}(Q_x^2 + Q_y^2) + c_{\perp} (Q_z-Q_{z0})^2$,
where $c_{\parallel}$ and $c_{\perp}$ are constants.

In the case of $k_{z0} \neq \pi/2c$ where 
the sizes of the cones become different, only a single cone 
of two electron Fermi cone nests
with one of two hole cones  
for $\pm {\bf Q}_1 \equiv$ (0, 0, $\pm 2k_{z0}$)
which is the nesting vector in this case; 
the other combination of cones are not nesting at the same time.
This situation decreases 
$\Pi_{12}({\bf Q}_1)$ in Eq. (\ref{eq:pi12}) mainly in the way 
that the term containing $\ln$ is divided into two terms 
each with half of the present coefficient; 
one half is not largely changed, keeping the ln$T$ singularity, 
and the other diminishes substantially 
since $T^2$ in the argument of $\ln$ is 
replaced by $T^2 + c_{\perp} (2\pi/c - Q_{z1} )^2$.

The magnitude of $\Pi_{12}({\bf Q})$ evaluated above
and its remarkable growth around $Q_z \cong 2k_{z0}$ are in a fair agreement with
Hase's estimation of $\Pi_{12}({\bf Q})$ based on 
FLAPW band calculations.\cite{rf:hase}
However, the prominent peaking with decreasing temperature is
moderated to a finite peak height of 0.265(eV)$^{-1}$. 
This discrepancy from the above obtained divergent behavior is 
due to 
the employed approximation 
linearizing $\varepsilon_{\pm}({\bf k})$ 
in $p_{\parallel}$ and $k_z-k_{z0}$. 
When $k_{z0}=\pi/2c$, the second order in $k_z-k_{z0}$ vanishes in 
$\varepsilon_{\pm}({\bf k})$ 
and so the linear approximation is not
so bad and Eq. (\ref{eq:pi12}) is valid. 
However, when $k_{z0} \neq \pi/2c$, the $(k_z-k_{z0})^2$ term in   
$\varepsilon_{\pm}({\bf k})$ 
suppresses the ln$T$ divergence, leaving a non-divergent peak
around $Q_z \sim 2k_{z0}$.
In Hase's result there is another peak at $Q_z = \pi/c$ 
presumably due to closeness to perfect nesting.
Its height is 0.27(eV)$^{-1}$,
slightly higher than the off-set peak.
In order to decide which is the highest 
it is necessary to carry out numerical work 
taking account of the dependences on parameters such as 
temperature, band filling and band parameters.
Anyway, $\Pi_{12}({\bf Q})$ is remarkably enhanced around
${\bf Q}_1$ due to partial nesting and around ${\bf Q}_0$
due to the closeness to perfect nesting.  \



As seen above, $\Pi_{12}({\bf Q})$ 
is very much enhanced around ${\bf Q}_1$ and ${\bf Q}_0$,
Eq. (\ref{eq:Kdiv}) is probable to diverge or to be greatly enhanced 
with decreasing temperature, 
if $U_0$ takes a realistic value of $\sim(2 \sim 4)$eV
since $K+U'=U_0$, from Eq. (\ref{eq:coupling}). 
This means a great enhancement of the amplitude
of the interband pair transfer process. 
This leads to a superconducting instability due to the two-band mechanism. 
The above divergence is the first to appear, 
among other possible divergences, with 
increasing $U_0$ or with decreasing temperature.

$T_c$ is determined as the temperature 
where the pair electron scattering diverges in both bands
with taking account of other 
interaction processes, as in Ref.\cite{rf:yamaji}  
Semi-quantitatively, however, taking account of the effective 
interband pair scattering coupling constant averaging 
$\Gamma_{12}$ in Eq. (\ref{eq:Kdiv}), 
by representing it by an effective coupling constant $\hat{K}$, 
and also of the intraband 
Coulomb coupling constant $U$, 
we obtain the mean-field-like 
superconducting gap equation with two gap parameters 
$\Delta_1$ and $\Delta_2$ for the two bands as in Ref.\cite{rf:yamajiabe} 
as follows:
\begin{eqnarray}
\Delta_i &=& \frac{1}{N} \sum_{{\mib k}}
[ W_i \theta(\omega_D - \vert \xi_{i{\mib k}}\vert) - U ] 
\frac{\Delta_i}{2E_{i{\mib k}}} 
\tanh \frac{E_{i{\mib k}}}{2k_B T} \nonumber \\
& & - \frac{1}{N} \sum_{{\mib k}}
\hat{K} \frac{\Delta_{\bar{i}} }{ 2E_{\bar{i}{\mib k}}} 
\tanh \frac{ E_{\bar{i}{\mib k}} }{ 2k_B T}, \quad
(i=1 \;  \mbox{or} \; 2) 
\label{eq:gapeq} 
\end{eqnarray}
where ${\bar i} = 3 - i$ and
$E_{i{\mib k}}=\sqrt{\xi_{i{\mib k}}^2 + \Delta_i^2}$; 
here we added an intraband phonon-mediated 
attractive coupling constant $W_i$ in band $i$ 
as a subsidiary ingredient
to consider about the isotope effect;  
$\xi_{i{\mib k}}$ is the energy of band $i$ with
reference to the chemical potential, 
and $\omega_D$ is the Debye energy. 
In a simple case where the geometrical means of the band half 
widths of two bands are equal, say to $D$, and the coupling 
characteristics are identical, the gap equation is easily solved to give 
\begin{eqnarray}
T_c &=& 1.13\omega_D\exp[-1/(\lambda - \mu^*)], \label{eq:Tc}  \\
\lambda &=& N_1(0)W_1,  \\
\mu^* &=& \frac{ N_1(0)(U-\hat{K})}{1+ N_1(0)(U-\hat{K})
\ln({D \over \omega_D})},  
\end{eqnarray}
where $N_1(0)$ is the state 
density per spin of band 1. In the present case $\hat{K}$ is 
enlarged due to the band nesting and the ladder process in 
Fig. \ref{fig:diagram}, $\mu^*$ becomes small and can be 
even negative and so $T_c$  can be very much elevated. 
It should be noted that the effect of $W_i$ is subsidiary;
even if $W_i=\lambda=0$, 
$T_c=1.13\omega_D\exp(-1/N_1(0)(\hat{K}-U))$ is finite if $\hat{K} > U$. 
One feature of this model is that $\Delta_1$ and $\Delta_2$ 
take the opposite signs although the gap is $s$-wave-like, 
being constant, in each band. 

The isotope exponent $\alpha$ is given by 
$\alpha=\{1-[\mu^*/(\lambda-\mu^*)]^2\}/2$. 
Applying this and Eq. (\ref{eq:Tc}) MgB$_2$ with $\alpha=0.26$ and 
$\omega_D=700 \mbox{K}$,\cite{rf:budko} 
we get $\mu^*=-0.23$ and $\lambda=0.103$. 
The latter value looks much smaller than the 
experimentally obtained value of $\lambda \sim 0.6$.\cite{rf:ott,rf:junod} 
However, the experimental $\lambda$ value 
extracted from the specific heat is 
an upper bound since the mass enhancement due to 
electron correlation is not taken into account.  
Further, it should be noted that our $\lambda$ is 
for a single $\pi$-band while the experimental $\lambda$ is 
the sum for all bands. 
So the above-mentioned value of $\lambda$ is not absurd. 
There is another point to note on the isotope effect, 
i.e., the increase of $T_c$ by 1K in Mg$^{10}$B$_2$ 
may have been brought about partly
through band changes and increase\cite{rf:hase} 
of $\Pi_{12}({\bf Q})$ due to expansion 
of lattice constant $a$ and slight decrease of $c$.\cite{rf:budko}

In the above consideration only the intraband phonon-mediated 
BCS attraction is taken into account with the interband BCS term neglected. 
The latter is in conflict with the $\hat{K}$ term. This neglect is justified 
since the electron scattering process from one $\pi$-band to the 
other by the in-plane stretching mode phonons was found to vanish 
due to symmetry of Bloch wave functions in the close neighborhood 
of the Brillouin zone boundary where important carriers exist. 

Any other combination in MgB$_2$, $e.g.$, between a 
$\pi$- and a $\sigma$-band, or between two $\sigma$-bands, 
there is no good nesting enhancing $\Pi_{12}$. Further the 
interband pair transfer coupling $K$ in these combinations 
is found to be one order of magnitude smaller than $U_0$. 
The largest contribution to $K$ among the $\sigma$-bands
was found to vanish due to symmetry in another tight-binding approximation.
So in the present framework, the role of the $\sigma$-bands 
are subsidiary with smaller gap parameters. 
However, there can be an opposite situation, unfortunately to us,
where the $\sigma$-bands are primary due to phonon-mediated 
attraction and the present mechansim of the $\pi$-bands are 
not important because the above derived superconducting instability
happens to be too weak
but it does not look natural in view of the jump-wise increase of $T_c$
of MgB$_c$ among diborides and the large value of $\Pi_{12}({\bf Q})$.
Recently Imada 
also presented a two-band consideration on 
MgB$_2$.\cite{rf:imada} He takes 
no account of nesting and the two bands in consideration 
are different from ours. \


In summary, 
coupling constant $K$ for the interband pair transfer process,
coupling constant $U'$ for interband Coulomb repulsion etc. 
in two $\pi$-bands of MgB$_2$ were
found to be given by half of the on-site Coulomb energy on the 
boron $p_z$-orbital, specifically in scattering processes 
relevant to the two-band mechanism. 
In terms of them the diagrammatic amplitude for the
interband pair scattering expressed by ladder diagrams 
is given by $K/[1-(K+U')\Pi_{12}({\bf Q})]$, where 
$\Pi_{12}({\bf Q})$ is the interband polarization function 
between two $\pi$-bands when wave vector ${\bf Q}$ is parallel to $\hat{z}$. 
Using the tight-binding bands for $\pi$-bands, 
$\Pi_{12}({\bf Q})$ was estimated and found to be quite large for 
${\bf Q}$ equal to 
${\bf Q}_1$ and ${\bf Q}_0$ which 
are parallel to $\hat{z}$ and
allow approximate nesting.
Therefore, $K/[1-(K+U')\Pi_{12}({\bf Q})]$ is very probale to be
divergently enlarged in MgB$_2$. This
necessarily brings about the two-band-type superconductivity 
whose $T_c$ can be very high. 
A semi-quantitative gap equation was given 
which takes also account of the BCS interaction in each band
as a subsidiary ingredient
and $T_c$ and the isotope exponent were derived.
Some characteristic features of the gap parameters for various bands
were briefly noted.
This mechanism was claimed to be a very probable candidate 
for the superconducting mechanism of MgB$_2$. 
The isotope exponent $\alpha = 0.26$ and
the electron-phonon $\lambda$ value $\sim$ 0.6,
taken by many researchers as supporting the
electron-phonon scenario, were pointed out not to be decisive.
The present mechanism gives a 
guiding principle to further enhance $T_c$ that 
the nesting between two $\pi$-bands should be improved.\

The author heartily thanks Dr. I. Hase, Professor J. Kondo, 
Dr. N. Shirakawa and Dr. T. Yanagisawa for useful information.

\end{document}